\documentclass[aps,prl,twocolumn,amsmath,amssymb, superscriptaddress,tightenlines]{revtex4}
\usepackage{graphicx}
\usepackage{epsfig}
\usepackage{dcolumn}
\usepackage{bm}
\usepackage{amssymb}
\usepackage{bm}
\usepackage{amsmath, bm}
\usepackage{upgreek}
\usepackage[colorlinks,citecolor=blue,linkcolor=blue,hyperindex]{hyperref}

\begin{document}

\title{Thermal interferometry of anyons in spin liquids}
\author{Zezhu Wei}
\affiliation{Department of Physics, Brown University, Providence, Rhode Island 02912, USA}
\affiliation{Brown Theoretical Physics Center, Brown University, Providence, Rhode Island 02912, USA}
\author{V. F. Mitrovi{\'c}}
\affiliation{Department of Physics, Brown University, Providence, Rhode Island 02912, USA}
\author{D. E. Feldman}
\affiliation{Department of Physics, Brown University, Providence, Rhode Island 02912, USA}
\affiliation{Brown Theoretical Physics Center, Brown University, Providence, Rhode Island 02912, USA}
\date{\today}


\begin{abstract}
Aharonov-Bohm interferometry is the most direct probe of anyonic statistics in the quantum Hall effect. The technique involves oscillations of the electric current as a function of the magnetic field  and is not applicable to Kitaev spin liquids and other systems without charged quasiparticles. 
Here, we establish a novel protocol, involving heat transport,  for  revealing  fractional statistics  even in the absence of charged excitations, as is the case in quantum spin liquids.   Specifically, we demonstrate that heat transport in Kitaev spin liquids through two distinct interferometer's geometries,  Fabry-Perot and Mach-Zehnder,
exhibits drastically  different  behaviors. Therefore, we propose the use of heat transport interferometry as a probe of  anyonic statistics  in charge insulators.
\end{abstract}

\maketitle


The last three years have seen important developments in probing fractional statistics in the quantum Hall effect \cite{review-FH}. One development was a Fabry-Perot interferometry experiment \cite{manfra20} at the filling factor $1/3$ (Fig. 1a). 
In the experiment, electric current flows through  two constrictions (QPC1 and QPC2).  The interference of the contributions from the two constrictions manifests itself in Aharonov-Bohm oscillations of the current in response to changing magnetic field.
The period of the oscillations is determined by the charge of the interfering quasiparticles. At some values of the field, the oscillation phase jumps. This happens because new anyons enter between the constrictions. The phase jumps encode the statistics of those anyons.
The physics is even more interesting for non-Abelian anyons in the second Landau level, where the even-odd effect is expected: as new anyons enter the device, an  interference picture alternatively turns on and off \cite{eo1,eo2,willett2019}. 

Anyons are electrically charged in the quantum Hall effect. Fractional statistics has also been long predicted in systems without charged excitations. Examples include the Kalmeyer-Laughlin \cite{semion2} and Kitaev \cite{kitaev-rev} spin liquids. A recent thermal conductance experiment \cite{RuCl3}
supports the presence of non-Abelian anyons in $\alpha-$RuCl$_3$, which is believed to host a Kitaev liquid \cite{RuCl3-K}. 
The interpretation of that experiment is currently debated \cite{phon-RuCl3-1,phon-RuCl3-2,czk}, 
and it is clear that new methods are needed to test quasiparticle statistics of neutral excitations. 
Interferometry is the most direct probe \cite{interferometry,amhma,kamda} of statistics since it involves 
running anyons around other anyons. However, the Aharonov-Bohm technique cannot work in the absence of charged quasiparticles. Thus, one can only implement it indirectly by conjugating a spin liquid with a system that can carry electric current \cite{amhma}. 

In this paper we show that a direct version of interferometry does not require charged excitations. It involves heat current instead of electric current and can be implemented in any systems since energy can flow in any system. The magnetic field ceases being a convenient experimental knob. Instead,
it becomes useful to compare transport in the Fabry-Perot \cite{interferometry} and Mach-Zehnder \cite{MZ-Heiblum,MZ1} geometries (Fig. 1).

 \begin{figure}[!htb]
\bigskip
\centering\scalebox{1}[1]
{\includegraphics{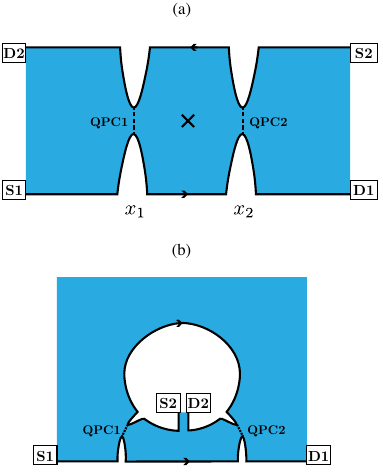}}
\caption{{Fabry-Perot (a) and Mach-Zehnder (b) interferometers. Heat travels from sources S1 and S2 to drains D1 and D2 along chiral edges and tunnels between the edges at the two point contacts shown with dashed lines. The cross shows a localized anyon.}}
\label{fig1:thermal}
\end{figure}

Multiple experiments on quantized thermal conductance in topological matter have been published in recent years. This includes work on the integer quantum Hall effect \cite{Jezouin}, 
the fractional quantum Hall effect in GaAs \cite{Banerjee2017,Banerjee2018} and graphene \cite{texp3,no-eq-graphene}, 
and the high-magnetic-field regime \cite{RuCl3,RuCl3-2} in $\alpha-$RuCl$_3$. At low temperatures, 
the gapped bulk of a topological material does not participate in heat conduction and only the edges matter. At the same time, any material contains gapless phonons, which should be taken into account in the interpretation of the data \cite{Jezouin,Banerjee2017,Banerjee2018}. Their interaction with the edges rapidly decreases as the temperature goes to zero 
\cite{phon-RuCl3-1,phon-RuCl3-2} and will be neglected below \cite{footnote0}. 
In the opposite limit of a strong interaction, interferometry cannot work due to the dominant dephasing of edge degrees of freedom by phonons \cite{footnote-temp}.

Our approach can be used with any fractional statistics. We will focus below on one particular anyon statistics predicted \cite{kitaev-rev} in a non-Abelian Kitaev spin liquid. Three types of excitations exist in  that liquid: trivial boson or vacuum $1$, Majorana fermion $\psi$, and Ising anyon $\sigma$. Any combination of quasiparticles belongs
to one of these sectors. Two quasiparticles can fuse in the following ways:

\begin{equation}
\label{eq-fusion}
\psi\times\psi=1;~\psi\times\sigma=\sigma; \sigma\times\sigma=1+\psi,
\end{equation}
where the final equality expresses two possible fusion outcomes for Ising anyons. The outcome of an interferometry experiment can be expressed in terms of the anyonic  topological spins \cite{kitaev-rev} $\theta_x$ and depends on the phase $\exp(i\phi^c_{ab})=\frac{\theta_c}
{\theta_a\theta_b}$ accumulated by anyon $a$ on a full counterclockwise circle around anyon $b$ under the assumption that the two anyons fuse to $c$. We will need the following phases \cite{MZ2}:

\begin{equation}
\label{eq-phases}
\phi^\sigma_{\sigma 1}=0;~ \phi^\sigma_{\sigma\psi}=\pi;~\phi^1_{\sigma\sigma}=-\pi/4;~\phi^\psi_{\sigma\sigma}=3\pi/4.
\end{equation}
Such phases are accumulated when one anyon $a=\sigma$ tunnels consecutively through the two point contacts in Fig. 1 while another anyon $b=1,\psi,\sigma$ 
is trapped between the two contacts.

Quantized thermal conductance is expected \cite{t1,t2,t3} when the two edges of a sample are far from each other. In interferometry, the edges are brought close in two points, and heat tunnels between the edges. This results in a non-universal correction to the thermal conductance. The correction depends on the tunneling amplitudes at the two contacts. 

 \begin{figure}[!htb]
\bigskip
\centering\scalebox{1.3}[1.3]
{\includegraphics{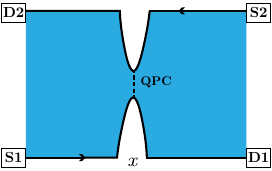}}
\caption{{A single tunneling contact is shown with a dashed line.}}
\label{fig2:thermal}
\end{figure}

We start with a brief discussion of a single contact (Fig. 2). The Lagrangian of the system in Fig. 2

\begin{equation}
\label{eq-1-contact}
L_1=L_{\rm edge,1}+L_{\rm edge,2}-T,
\end{equation}
where $L_{{\rm edge,}n}$ are the Lagrangians of the two edges and $T$ describes tunneling. In the simplest model, the edges host free Majorana fermions \cite{kitaev-rev}:

\begin{equation}
\label{eq-edge}
L_{{\rm edge,}n}=i\int dx\psi_n(\partial_t\pm v\partial_x)\psi_n,
\end{equation}
where $v$ is the edge velocity and the sign shows the propagation direction. 
The tunneling operator creates two excitations that fuse to vacuum on the opposite sides of the contact. In the Ising order each particle is its own antiparticle and hence $T$ creates two quasiparticles of the same type.
In principle, $T$ describes tunneling of multiple quasiparticle types and  includes an infinite number of perturbations to the sum of the edge actions (\ref{eq-edge}). At low temperatures and weak tunneling, the only perturbations that matter are relevant in the renormalization group sense. Our theory allows only one such perturbation

\begin{equation}
\label{eq-T}
T=\exp(-i\pi/16)\Gamma\sigma_2(x)\sigma_1(x),
\end{equation}
where $\sigma_2$ and $\sigma_1$ create Ising anyons in point $x$ on the upper and lower edges respectively, $\Gamma>0$ is the tunneling amplitude \cite{footnote1}, and 
the exponential factor \cite{amhma,sigma-tun}
ensures hermiticity and equals $1/\sqrt{\rm topological ~spin}$. 

We are interested in the deviation of the thermal current between the terminals in Fig. 2 from the quantized value.
The deviation equals the tunneling thermal current $I_T$  through the point contact between the two edges. The current depends on the temperature difference between two 
sources S1 and S2
and hence the two edges that emanate from the source terminals. It also depends on the tunneling amplitude $\Gamma$ and scales as $\Gamma^2$ in the lowest order of the perturbation theory at small $\Gamma$,
$I_T=r(T_1,T_2)\Gamma^2$, where $T_1$ and $T_2$ are the temperatures of the two edges, and the factor $r(T_1,T_2)$ depends on details of the edge physics and can be computed with Fermi's golden rule:

\begin{equation}
\label{eq-Fermi-r}
r=\frac{2\pi}{\hbar}\sum_{mn}\Delta E |\langle m|T/\Gamma |n \rangle|^2 \delta(E_m-E_n) P_n(T_1,T_2),
\end{equation}
where $|n\rangle, |m\rangle$ are eigenstates of the edge Hamiltonian,  $E_{n,m}$ are the combined energies of the two edges in those states,  $P_n$ is the Gibbs distribution, and $\Delta E$ is the energy change of the upper edge in the $|n\rangle\rightarrow|m\rangle$ process. 

The perturbative calculation  is applicable as long as the tunneling heat current is much smaller than the total heat current along the edges. In the simplest model (\ref{eq-edge}),
$r\sim T_1^{1/4}$ at a constant ratio $T_1/T_2$ since the scaling dimension \cite{kitaev-rev} of $\sigma$ is $1/16$. 

We turn to a Fabry-Perot interferometer (Fig. 1a) now. The Lagrangian differs from (\ref{eq-1-contact}) only by the presence of two tunneling terms in $T$:

\begin{eqnarray}
\label{eq-T-FP}
T=\exp(-i\pi/16)\Gamma_1\sigma_2(x_1)\sigma_1(x_1)\nonumber\\
+\exp(-i\pi/16)\Gamma_2\sigma_2(x_2)\sigma_1(x_2),
\end{eqnarray}
where $x_1$ and $x_2$ are the locations of the two point contacts. The tunneling amplitudes $\Gamma_1$ and $\Gamma_2$ can be determined  experimentally up to the factor $r(T_1,T_2)$ from comparison with a single-point-contact geometry. 
For such a comparison, one needs to fabricate a single point contact in exactly the same way as one of the two contacts in the interferometer and measure the tunneling heat current.

We will assume that the thermal length is much longer than the distance between the two QPCs, $hv/k_BT\gg x_2-x_1$. This will allow us to treat the interferometer as a single point contact in an effective low-energy theory for energies $E\sim T_{1,2}$. In the absence of trapped quasiparticles, the low-temperature physics of the interferometer then reduces to that of 
a tunneling contact with the tunneling amplitude $\Gamma=\Gamma_1+\Gamma_2$.
Thus, the thermal current through the interferometer that contains no trapped topological charge

\begin{equation}
\label{eq-FP-1}
I_{{\rm FP}, 1}=r(T_1,T_2) (\Gamma_1+\Gamma_2)^2,
\end{equation}
where the factor $r(T_1,T_2)$ is the same as for a single point contact.
The above expression can be understood as the result of constructive interference of the two paths from the lower edge to the upper edge via the two point contacts.
 
A trapped Majorana fermion contributes a phase of $\pi$ to any trajectory that encircles it.  Hence, when the interferometer contains the topological charge $\psi$, there is a phase difference of $\pi$ for the trajectories via the two point contacts. Interference becomes destructive:

\begin{equation}
\label{eq-FP-psi}
I_{{\rm FP}, \psi}=r(T_1,T_2) (\Gamma_1-\Gamma_2)^2.
\end{equation}

A particularly interesting situation presents itself if the trapped topological charge in the device is $b=\sigma$. We need to consider separately two possibilities for the fusion channel of the tunneling and trapped anyons, $c=1$ and $c=\psi$. The processes in the two fusion channels do not interfere with each other just like interference is absent for 
electron transport in the two spin channels. The probabilities of the two channels are equal as follows from the general expression \cite{MZ-noise}

\begin{equation}
\label{eq-fusion-p}
P^c_{ab}=N^c_{ab}\frac{d_c}{d_a d_b},
\end{equation}
where the fusion multiplicities $N^1_{\sigma\sigma}=N^\psi_{\sigma\sigma}=1$ and the quantum dimensions $d_1=d_\psi=1,~d_\sigma=\sqrt{2}$.  According to Eq. (\ref{eq-phases}), the interference phases in the two channels differ by $\pi$. Thus,

\begin{eqnarray}
\label{eq-FP-sigma}
I_{{\rm FP},\sigma}=\frac{r(T_1,T_2)}{2}\times\nonumber\\
\left[|\Gamma_1+\exp(-i\pi/4)\Gamma_2|^2
+|\Gamma_1-\exp(-i\pi/4)\Gamma_2|^2\right]\nonumber\\
=r(T_1,T_2)[\Gamma_1^2+\Gamma_2^2].
\end{eqnarray}

If one can control the topological charge of the interferometer, the observation of the three regimes  (\ref{eq-FP-1},\ref{eq-FP-psi},\ref{eq-FP-sigma}) would prove the Ising anyonic statistics. At present it is unclear \cite{amhma} how to control the trapped charge, 
so it may happen that every interferometer is always in the same regime. This can only happen \cite{footnote-2} in the regime (\ref{eq-FP-1}), in which case the experiment is not very informative: the behavior (\ref{eq-FP-1}) can be observed for any statistics, if the trapped topological charge is trivial.
We thus turn to Mach-Zehnder interferometry (Fig. 1b) that does not require control of
the trapped topological charge.

 \begin{figure}[!htb]
\bigskip
\centering\scalebox{1.3}[1.3]
{\includegraphics{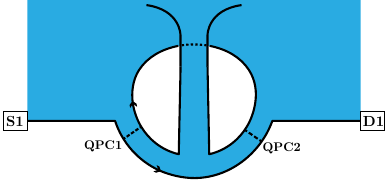}}
\caption{{The Mach-Zehnder setup is topologically equivalent to a setup with an infinite open inner edge.}}
\label{fig3:thermal}
\end{figure}

In Mach-Zehnder interferometry, the topological charge inside the device changes after each tunneling event \cite{MZ1,review-FH}. Indeed, each anyon that tunnels into the inner edge of the device is eventually absorbed by a drain located inside the device. Fig. 3 depicts an alternative setting: 
a tunneling anyon ends up on a very long edge, which is inside the interference loop topologically. As a consequence, the tunneling probability changes for each tunneling event. We thus need to introduce a family of tunneling rates $p^c_{\sigma b}$, where $b$ is the trapped topological charge and
$c$ is the fusion channel of $b$ and the tunneling anyon $\sigma$. As before, we assume that the distance between the point contacts along each edge is much shorter than the thermal length. 
The tunneling Hamiltonian $\hat T=\Gamma_1\hat T_1+\Gamma_2\hat T_2 e^{i\alpha}$ contains two real amplitudes $\Gamma_{1,2}$, two operators $\hat T_{1,2}$ that transfer an anyon from the outer edge to the inner edge, and a phase $\alpha$, which ensures hermiticity. We will find $\alpha$ from the condition that the same tunneling rates obtain 
from $\hat T$ and the tunneling Hamiltonian $\hat T^\dagger=\hat T=\Gamma_1 \hat T_1^\dagger+\Gamma_2\hat T_2^\dagger e^{-i\alpha}$, where $\hat T_{1,2}^\dagger$ transfer anyons from the inner edge to the outer edge.

We first use $\hat T$. Only one fusion channel exists for $b=1,\psi$, and one finds

\begin{eqnarray}
\label{eq-p-1}
p^\sigma_{\sigma 1}=p(T_1,T_2)|\Gamma_1+\Gamma_2 e^{i\alpha}|^2;\\
\label{eq-p-psi}
p^\sigma_{\sigma\psi}=p(T_1,T_2)|\Gamma_1-\Gamma_2 e^{i\alpha}|^2,
\end{eqnarray}
where the expression for $p(T_1,T_2)$ follows from Fermi's golden rule and is similar to (\ref{eq-Fermi-r}):

\begin{equation}
\label{eq-Fermi-p}
p=\frac{2\pi}{\hbar}\sum_{mn} |\langle m|\sigma_2(x)\sigma_1(x) |n \rangle|^2 \delta(E_m-E_n) P_n(T_1,T_2).
\end{equation}

Two fusion outcomes are possible for $b=\sigma$. The probabilities of the fusion outcomes are identical, but the tunneling rates for the two outcomes are not:

\begin{eqnarray}
\label{eq-p-sigma-1}
p^1_{\sigma\sigma}=\frac{p(T_1,T_2)}{2}\left|\Gamma_1+\Gamma_2\exp
(i\alpha-i\pi/4)\right|^2; \\
\label{eq-p-sigma-psi}
p^\psi_{\sigma\sigma}=\frac{p(T_1,T_2)}{2}\left|\Gamma_1-\Gamma_2\exp(i\alpha-i\pi/4)\right|^2.
\end{eqnarray}
The total tunneling rate

\begin{equation}
\label{eq-p-sigma}
p_{\sigma\sigma}=p(T_1, T_2)[\Gamma_1^2+\Gamma_2^2].
\end{equation}
Repeating the same calculation \cite{sup} with $T^\dagger$ gives the same set of answers for $\alpha=\pi/8$, which is thus the right choice in the above equations.

 \begin{figure}[!htb]
\bigskip
\centering\scalebox{1}[1]
{\includegraphics{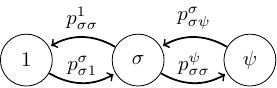}}
\caption{{The symbols in the circles show the trapped topological charge in a Mach-Zehnder interferometer. Arrows show possible transitions.}}
\label{fig4:thermal}
\end{figure}

The average energy transferred between the edges in each tunneling event is the same\cite{footnote3}: $\overline{\Delta E}=r(T_1,T_2)/p(T_1,T_2)$. Thus, to compute the heat current we need to find the average number of tunneling events per unit time.
All possible tunneling events are represented by the diagram in Fig. 4. The states of the interferometer are labeled with the trapped topological charge $b$. If we assume that the initial trapped charge is $b=\sigma$, the average time until a tunneling event
$t_\sigma=1/p_{\sigma\sigma}$. The probabilities of tunneling into states with $b=1$ and $b=\psi$ are $q_1=p^1_{\sigma\sigma}/p_{\sigma\sigma}$ and $q_\psi=p^\psi_{\sigma\sigma}/p_{\sigma\sigma}$ respectively. The average times until tunneling from the 
states with  $b=1$ and $b=\psi$ to $b=\sigma$ are $t_1=1/p_{\sigma 1}$ and $t_\psi=1/p_{\sigma\psi}$. After two tunneling events the system always returns to $b=\sigma$. The average time of two tunneling events is given by

\begin{equation}
\label{eq-bar-t}
\bar t=t_\sigma+q_1t_1+q_\psi t_\psi.
\end{equation}
Hence, the thermal current becomes

\begin{equation}
\label{eq-I-MZ}
I_T=\frac{2\overline{\Delta E}}{\bar t}=r(T_1,T_2)(\Gamma_1^2+\Gamma_2^2).
\end{equation}
For comparison, if the tunneling particles are bosons or fermions, the behavior is the same as in the Fabry-Perot setup, Eq. (\ref{eq-FP-1}). 
Thus, the observation of the contrasting behavior (\ref{eq-FP-1}) and (\ref{eq-I-MZ}) in the two setups is a signature of fractional statistics is a system without charged quasiparticles.

The results for a Mach-Zehnder interferometer do not change if topological charge can tunnel between the edges of the device and localized states in the bulk as long as the typical time between tunneling events exceeds the time between tunneling events at the point contacts.  This is not the case in the Fabry-Perot setup, where the tunneling to localized states 
must be slow on the laboratory time-scale to have no effect on the current. Rare tunneling events into localized states in the Fabry-Perot geometry lead to strong telegraph noise \cite{kane-tel} that serves as another signature of fractional statistics. In the absence of tunneling into localized states, telegraph noise can be induced by making a hole in the device 
as discussed in Supplemental Material \cite{sup}.

{\color{black} To measure edge heat currents, one can transfer all or a fraction of the heat current to a quantum Hall (QH) edge. The transferred heat current can be extracted from the temperature of a piece of metal, connected to a QH edge \cite{sup,Jezouin,Banerjee2017}.  
One approach to heat transfer between a Kitaev magnet and a QH system relies on a superconductor as an intermediary \cite{amhma}. Alternatively, one can transfer heat from the spin liquid to a conductor connected with a QH edge. 
Fig. 5 illustrates the setup for a quantum wire in the role of the conductor. A hole is created in the Kitaev liquid near the edge so that anyon tunneling is possible between the edge of the hole and the outer edge of the liquid in the constriction.
 The most relevant interaction of the wire and the Kitaev magnet is 
$W=\sigma_h\sigma_e\partial_x\phi$, where $\sigma_{h,e}$ create Ising anyons on the edge of the hole and the outer edge of the liquid; $\partial_x\phi$ is proportional to the charge density in the wire. Near a resonance, the Hamiltonian of the Kitaev liquid contains no tunneling term $\sigma_h\sigma_e$. The scaling dimension of $W$ is $9/8$ and $W$ ensures 
significant energy exchange between the wire and the magnet.

 \begin{figure}[!htb]
\bigskip
\centering\scalebox{1}[1]
{\includegraphics{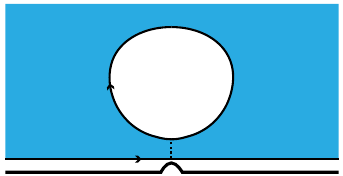}}
\caption{{A dashed line shows a constriction between the outer edge and the edge of the hole in a Kitaev magnet. The thick line at the bottom represents a conductor in contact with a magnet edge and a QH edge (not shown).}}
\label{fig05:thermal}
\end{figure}}

In conclusion, interferometry allows probing fractional statistics with heat transport. 
Information about heat currents can be extracted from the temperatures of the source and drain reservoirs. 
The temperatures of electrically conducting  reservoirs can be found from noise  \cite{Jezouin,Banerjee2017,Banerjee2018} or quantum dot \cite{qd1,qd2,qd3,qd4,qd} thermometry.
The contrasting behavior of Mach-Zehnder and Fabry-Perot interferometers is a smoking gun evidence of fractional statistics.

\section*{Acknowledgments} DEF and ZW were supported in part by the NSF under grant No. DMR-1902356.  
VFM was supported in part by the NSF under grants  QLCI-1936854 and DMR-1905532. 

{\it Note added:} After the completion of this paper we became aware of Ref. \cite{added} that overlaps with our work.

\appendix

\section{Appendix}

\subsection{Tunneling in the Mach-Zehnder geometry}

The goal of this section consists in finding the phase $\alpha$ in the tunneling Hamiltonian $\hat T=\Gamma_1\hat T_1+\Gamma_2\hat T_2 e^{i\alpha}$.
We will follow two approaches: the second approach compares $\hat T$ and $\hat T^\dagger$  which must be the same due to the hermiticity of the Hamiltonian. The first approach
uses detailed balance and bypasses the use of $\hat T^\dagger$.

 \begin{figure}[!htb]
\bigskip
\centering\scalebox{1.3}[1.3]
{\includegraphics{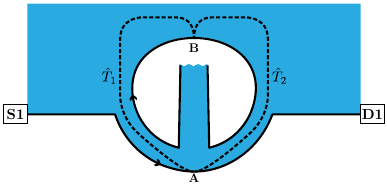}}
\caption{{In the low-energy effective model of a Mach-Zehnder interferometer, the two tunneling operators describe anyon transfer along the two dashed lines.}}
\label{fig6:thermal}
\end{figure}

In the first approach, we assume that the temperatures $T_1=T_2=T$. There is no heat current in that case. Yet, the tunneling probabilities 
(12-16)
 are nonzero [all equation numbers refer to the main text]. The detailed balance principle states that the probability of the transition from the trapped topological charge $a$ to the trapped charge $b$ is related to the probability of the reverse transition $b\rightarrow a$. Indeed, the two probabilities are equal to
 
 \begin{equation}
\label{eq-Fermi-p-a-b}
p^b_{\sigma a}=\frac{2\pi}{\hbar}\sum_{mn}|\langle m |\hat T |n \rangle|^2 \delta(E_m-E_n) P^a_n(T,T),
\end{equation} 
\begin{equation}
\label{eq-Fermi-p-b-a}
p^a_{\sigma b}=\frac{2\pi}{\hbar}\sum_{mn}|\langle n |\hat T |m \rangle|^2 \delta(E_m-E_n) P^b_m(T,T)
\end{equation} 
and differ only by the partition-function denominators in the Gibbs factors  $P^a_n(T,T)=\frac{\exp(-E_n/T)}{Z_a(T)}$ and $P^b_m(T,T)=\frac{\exp(-E_m/T)}{Z_b(T)}$, where $E_n=E_m$ and the two partition functions are computed in two superselection sectors. The partition functions do not depend on $\Gamma_{1,2}$. Hence,
 the ratio of the $a\rightarrow b$ and $b\rightarrow a$ transition probabilities does not depend on $\Gamma_{1,2}$.  The transition probabilities are given by equations 
 (12,13,15,16).
 $\alpha$ can be found from their comparison.
 
 The second approach starts with an effective low-energy model in which the two tunneling operators $\hat T_{1,2}$ move anyons between the same points $A$ and $B$ on the lower and upper edges. It is illustrated in Fig. 6. Let $\Pi_a=\Pi^\dagger_a$ be the projector to the sector with the trapped topological charge $a$. We observe that
 
 \begin{equation}
 \label{eq-t1-t2}
 \Pi_c \hat T_2 \Pi_b=\Pi_c \hat T_1 \Pi_b\exp(i\phi^c_{\sigma b}).
 \end{equation}
 We now use the hermiticity of $\exp(i\alpha)\hat T_2=\exp(-i\alpha)\hat T^\dagger_2$:
 
 \begin{equation}
 \label{eq-t-tdagger}
 \exp(i\alpha)\Pi_c \hat T_2 \Pi_b = \exp(-i\alpha)(\Pi_b \hat T_2 \Pi_c)^\dagger
 \end{equation}
 and hence
 
 \begin{equation}
 \label{eq-t-tdagger-1}
 \exp(i\alpha)\frac{\theta_c}{\theta_\sigma\theta_b}\Pi_c \hat T_1 \Pi_b = \exp(-i\alpha)\frac{\theta_c\theta_\sigma}{\theta_b}\Pi_c \hat T_1 \Pi_b.
 \end{equation}
 It follows that $\exp(i\alpha)=\pm\theta_\sigma=\pm\exp(i\pi/8)$. The choice of the sign does not affect the physics.
 
 \subsection{Telegraph noise}

 \begin{figure}[!htb]
\bigskip
\centering\scalebox{1.3}[1.3]
{\includegraphics{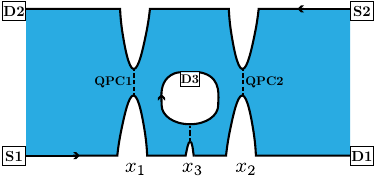}}
\caption{{Fabry-Perot interferometer with a hole. Topological charge can tunnel to the edge of the hole via a tunneling contact at point $x_3$.}}
\label{fig5:thermal}
\end{figure}

Assume that a closed inner edge is present in a Fabry-Perot device and that weak tunneling occurs between that edge and lower edge $1$ in point $x_3$ as shown in Fig. 7. We will assume that the inner edge is connected to a drain so that its temperature remains equal to the ambient temperature. In that case, the probability of a tunneling event in point $x_3$
does not depend on the history. Further, assume that the average time $\tau$ between tunneling events in point $x_3$ is large: $\tau\gg 1/[p(T_1,T_2)\Gamma_{1,2}^2]$. The topological charge $b$ localized on the inner edge changes according to the arrows in Fig. 4 of the main text, but the tunneling rates are the same and equal to $1/\tau$ in all nodes of the diagram. 
Thus, $b=\sigma$ half of the time, $b=1$ quarter of the time, and $b=\psi$ quarter of the time. The heat current between the upper and lower edges depends on $b$ according to equations 
(11,8,9).
The average heat current 
$I_T$ is the same as at $b=\sigma$, Eq. 
(11).
At $b=1,\psi$, the current deviates from the average by $\pm 2r(T_1,T_2)\Gamma_1\Gamma_2$. We can now compute the low-frequency noise of the heat current

\begin{eqnarray}
\label{eq-noise}
S=\lim_{\omega\rightarrow 0}\int_{-\infty}^{+\infty} dt \langle I_T(0) I_T(t)+I_T(t) I_T(0)\rangle\exp(i\omega t)\nonumber\\
=8\tau r^2(T_1,T_2)\Gamma_1^2\Gamma_2^2.~~
\end{eqnarray}
One expects high noise at large $\tau$.
 
 \subsection{Measurement of the heat current}


 \begin{figure}[!htb]
\bigskip
\centering\scalebox{1.3}[1.3]
{\includegraphics{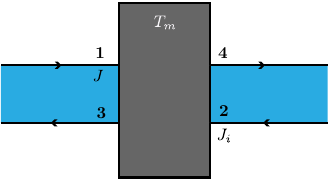}}
\caption{{Chiral quantum Hall edges 1 and 2 enter a piece of metal, and channels 3 and 4 exit it.}}
\label{fig7:thermal}
\end{figure}

We assume that the heat current $J$ is carried from the contact point of Kitaev and quantum Hall (QH) edges by the QH edge that exchanges heat with the magnetic material. We label that QH edge as edge 1. The heat current $J$ can be measured with a high precision with the method from Refs. [18,19] of the main text.
Edge 1 is absorbed by a small piece of metal (Fig. 8) which also absorbs edge 2 in equilibrium at the ambient temperature $T_0$. Edge 2 carries the heat current $J_i=KT_0^2/2$ into the piece of metal, where $K$ is the quantized thermal conductance of a QH edge.
Two chiral edges (3 and 4) emanate from the piece of metal. Their temperatures equal the temperature $T_m$ of the piece of metal. Hence, they carry the total heat current $J_o=2KT_m^2/2$. From the energy conservation, $J=J_o-J_i$. Hence, $J$ can be found from the known temperature $T_0$
and the temperature $T_m$. The latter temperature can be determined from the low-frequency electric-current noise on edge 3. The noise is defined as

\begin{equation}
\label{eq-S3}
S_3=\lim_{\omega\rightarrow 0}\int_{-\infty}^\infty dt \langle I_3(t)I_3(0)+I_3(0)I_3(t)\rangle\exp(i\omega t),
\end{equation}
where $I_3$ is the electric current. The noises on the other edges have similar definitions. According to the fluctuation-dissipation theorem, the noise $S_2=2GT_0$, where $G$ is the quantized electrical conductance of a QH edge.

The incoming currents $I_1$ and $I_2$ are uncorrelated. This does not hold for the currents $I_3$ and $I_4$. The currents
$I_{3,4}$ can be represented as

\begin{equation}
\label{eq-I3}
I_n=I^T_n+G\delta V,
\end{equation}
where $I^T_n$ is the ``Nyquist'' contribution which alone would produce the noise $2GT_m$. It describes thermal fluctuations and results from the Gibbs distribution for the outgoing electrons.
The second contribution is due to the voltage fluctuations $\delta V$ in the piece of metal. The voltage fluctuations can be found from charge conservation:

\begin{equation}
\label{eq-delta-V}
2G \delta V +I^T_3+I^T_4=I_1+I_2.
\end{equation}
Since the currents $I_1$, $I_2$, $I^T_3$, and $I^T_4$ are uncorrelated, computing $S_3$ is straightforward. An amplifier, measuring that noise, is always connected with another chiral channel at the ambient temperature $T_0$. 
Thus, the observed noise

\begin{equation}
\label{eq-answer-S}
S=GT_m+\frac{S_1}{4}+\frac{5}{2}GT_0,
\end{equation}
where $S_1$ can be measured separately.

Equation (\ref{eq-answer-S}) allows computing the heat current $J$. In general, this heat current differs from the heat current $J_M$ 
flowing from the interferometer along the edge of a Kitaev magnet. Fortunately, the absolute value of $J_M$ is of little interest. 
In the presence of anyons in the interferometer, the heat currents $J_M$ and $J$ acquire corrections, proportional to $\Gamma_{1,2}^2$. Information about fractional statistics can be extracted from the comparison of the corrections in different configurations.


\begin{thebibliography}{99}
\bibitem{footnote0} 
{\color{black} In $\alpha-$RuCl$_3$, the edge-phonon relaxation length is much greater \cite{phon-RuCl3-2} than the lattice spacing $a\sim 1~{\rm nm}$ at $T\sim 5~{\rm K}$. 
It diverges as a high negative power of the temperature and becomes very large at $T\sim 0.5~{\rm K}$. }
\bibitem{footnote-temp}
{\color{black}The observed quantization of the thermal conductance in the experiment \cite{RuCl3} is a consequence of the strong Majorana-phonon coupling at high temperatures \cite{phon-RuCl3-1,phon-RuCl3-2}. No quantization is expected at low temperatures, where the coupling is weak. 
In particular, probes of the lattice temperature, in the case of weak coupling, would yield a much lower thermal conductance than its quantized value, observed in the strong coupling limit. In our approach, thermal probes involve electronic degrees of freedom, such as QH edge modes, and do not involve phonons. Moreover, strong coupling with
 acoustic phonons wipes out the universal behavior. Hence, here universal results are expected at low temperatures, where the Majorana-phonon interaction is suppressed.}
\bibitem{footnote1} A negative $\Gamma$ is also consistent with hermiticity, but can be interpreted as a positive $\Gamma$ in the presence of a Majorana fermion in the bulk.
\bibitem{footnote3} {\color{black} The functions $r(T_1,T_2)$ are approximately equal in the Mach-Zehnder and Fabry-Perot setups, provided that the interferometer size is much greater than the lattice spacing $a\sim 1~{\rm nm}$.  The same condition allows neglecting finite-size corrections to the statistical phase.}
\end{thebibliography}
\end{document}